\documentclass[11pt,twoside]{article}
              
\usepackage{asp2004}
\usepackage{epsf}
\usepackage{psfig}
\usepackage{lscape}
\usepackage{graphicx}

\begin{document}
\title{Observations of V838 Mon and the nearby region in the \boldmath
  CO $J$ = 1$\to$0, 2$\to$1 and 3$\to$2 transitions}   %%% Fill in title
\author{Tomasz Kami\'{n}ski$^1$, Martin Miller$^2$, Ryszard
  Szczerba$^3$, Romuald Tylenda$^{1,3}$}  
\affil{$^1$Nicolaus Copernicus University, Gagarina~11,
  87-100 Toru\'{n}, Poland; 
  $^2$I.~Physikalisches Institut, Universit\"{a}t zu K\"{o}ln,
  Z\"{u}lpicher Strasse~77, 50937 K\"{o}ln, Germany; $^3$Department
  for Astrophysics, Nicolaus Copernicus
  Astonomical Center, Rabia\'{n}ska~8, 87-100 Toru\'{n}, Poland}

\begin{abstract} %%% Abstract to run on from here.
We present observations of V838~Mon and its close
vicinity in the three lowest rotational transitions of CO.
The $J$
= 2$\to$1 and 3$\to$2 data were obtained using the 3 m KOSMA telescope. They
include on-the-fly maps covering a large area ($\sim$3.4 sq. deg.) 
around V838~Mon and long integrations on the star
position. Complementary observations in the CO $J$ = 1$\to$0
transition were obtained using the 13.7 m Delingha telescope. 
The star position as well as 25 other points preselected in the near vicinity 
of the object have been measured in this transition. 

We report on a detection of two narrow emission components in $J$ =
2$\to$1 and 3$\to$2 transitions at the position of V838~Mon. Lines were found at
radial velocities of $V_{\rm lsr}=53.3$~km~s$^{-1}$ and $V_{\rm
  lsr}=-11.0$~km~s$^{-1}$. Their origin is unclear. We also
shortly discuss results of the observations of the vicinity of V838~Mon.    
\end{abstract}

\section{Introduction}
The enigmatic eruption of V838 Mon, followed by its spectacular light echo,
triggered research in different fields of astrophysics. Apart from
studies of the evolution of the object, investigations of
the circumstellar and interstellar neighborhood of the star can also
be important for understanding the nature of the event.
We present observations undertaken to search for molecular matter
in the vicinity of V838 Mon.          

Using the results of the CO $J$~=~1$\to$0 galactic survey of \citet{dame}
\citet{loon} have suggested that V838 Mon is situated in a bubble of CO emission
of a diameter of about 1$^{\circ}$ (see Fig. 1). 
According to these authors the structure is of circumstellar origin due to
the AGB activity of the V838 Mon progenitor. Several critical points against this
finding and interpretation have been raised in
\citet{tyl}. However, the structure 
in the CO map reported in \citet{loon} is quite suggestive 
and we have found important to
obtain CO observations of the same ragion with a better sensitivity 
and angular resolution

There are also other important reasons for observing the star and
its vicinity in molecular lines. 
These are the nature of the echoing matter and a search for matter
lost during and after the 2002 outburst. The detection and
monitoring of the SiO maser emission from V838~Mon 
\citep{deguchi,claus} shows that a molecular
activity started close to the star. Complementary observations in molecular
thermal transitions would be important to better understand what is going on there.

\begin{figure}
\includegraphics[scale=0.6, origin=r, trim = 0 0 0 100]{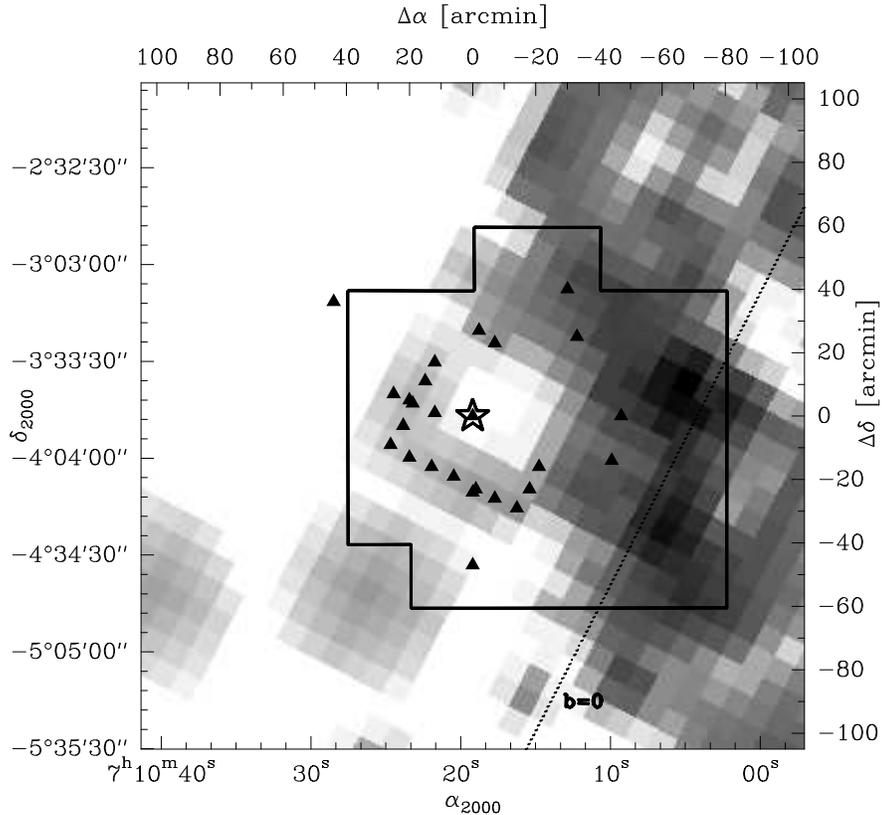}
\caption{A $3^{\circ}.5\times3^{\circ}.5$ map centered at the V838~Mon
position from the CO~(1--0) galactic survey \citep[data taken from
  {\it SkyView} -- {\tt \small
    http://skyview.gsfc.nasa.gov/};][]{dame}. The star symbol marks V838 
Mon position. The polygon drawn with a solid line defines the region
observed in the on-the-fly mode in the CO (2--1) and (3--2) transitions. Triangles mark
the positions observed in the CO (1--0) transition. Dashed line indicates the
Galactic equator.}
\end{figure}    
%%%%%%%%%%%%%%%%%%%%%%%%%%%%%%%%%%%%%%%%%%%%%%%%%%%%%%%%%%%%%%%%%%%%%%%    
\section{Observations}
We performed observations in the three lowest rotational transitions of CO
in millimeter and submillimeter wavelengths. Observations in CO
$J$~=~1$\to$0 (115.27 GHz) were obtained using the Delingha 13.7~m
telescope in October--November 2005. Beside the star position, 25 points in
the region around V838 Mon were observed, with majority of them
located in the molecular ring claimed in \citet{loon}. All the points
observed in CO (1--0) are marked with filled
triangles in Fig.~1.    

Observations in the higher transitions, i.e. CO $J$~=~2$\to$1 and 3$\to$2
(230.54 and 345.80 GHz, respectively), were obtained with the 3~m
KOSMA telescope. In April 2005 we made two {\it on-the-fly} maps with
a 1 arcmin spacing and of a total area of 3.4 squared deg. Location of
the mapped region is shown on Fig.~1. We also made longer integrations on
the V838~Mon position in April and December 2005 in both transitions.

In all the cases heterodyne SIS receivers and acousto--optical spectrographs
were used as frontends and backends, respectively.
%%%%%%%%%%%%%%%%%%%%%%%%%%%%%%%%%%%%%%%%%%%%%%%%%%%%%%%%%%%%%%%%%%%%%%%
\section{CO regions around V838 Mon}
In this section we present observations of the field around V838 Mon in the
three transitions, starting from those obtained with the KOSMA
telescope and followed by a discussion of the lowest transition observed
with the Delingha telescope.  
\subsection{On-the-fly maps in CO \boldmath $J$ = 2$\to$1 and 3$\to$2}
As a result of the on-the-fly observations we have obtained 
a set of 12~400 spectra for each
transition with typical integration time of 4--8 s. Basic technical
description of the data is given in Table~1. It should be noted that
the sensitivity of the survey in both lines is rather poor.
Furthermore, because of changing atmospheric
conditions the data
quality is not homogeneous. Hence, special data reduction procedures were needed to
visualize the data (e.g. clipping method). 

From an analysis of the spectra averaged over all the positions and from the channel
maps we have found that the molecular emission in the mapped 
region comes mainly from matter at radial velocities of $V_{\rm lsr}$=18--32 km~s$^{-1}$ and $V_{\rm
  lsr}$=44--57~km~s$^{-1}$. Adopting a galactic rotation curve from
\citet{rot} the two velocity ranges correspond to kinematical distances of $d_{\rm kin}$=2.3--3.0 kpc and $d_{\rm
  kin}$=6.2--6.5 kpc, respectively. These values are in good agreement
with distances to two spiral arms in the direction of V838~Mon,
i.e. Perseus Arm (3 kpc) and Norma--Cygnus Arm (6.25 kpc). We have
estimated physical parameters for the clouds emitting in
CO. Adopting $d_{\rm kin}$ as distances to these structures we have got
values of the physical parameters typical for molecular clouds.

In Fig.~2 we present the map of intensity integrated over
two velocity ranges (spectra used to obtain map were earlier clipped at
$3\sigma_{\rm rms}$ level) for the CO~(2--1) transition. The CO~(3--2)
map looks similar so we do not present it here. No significant
emission can be seen around V838 Mon up to a radius of about 40'.
In particular there is no shell-like structure around the star, 
contrary to the claim of \citet{loon}. All the
significant emission west from the star position comes from ordinary
giant molecular clouds in the Galactic disk.   
\begin{table}
\caption{Characteristics of the observations of the region around V838~Mon
  in the three lowest rotational transitions of the CO molecule.}
\smallskip
\begin{center}
{\small 
\begin{tabular}{lccc}
\tableline
\noalign{\smallskip}
CO line& $J$ = 1$\to$0&  $J$ = 2$\to$1&  $J$ = 3$\to$2\\
\noalign{\smallskip}
\tableline
\noalign{\smallskip}
beam&  55''& 130"&     82"\\
$T_{\rm sys}$ [K]&~~231 $\div$ 286&~~153 $\div$ 230&~~230 $\div$ 377\\
vel. range [km s$^{-1}$] &--139 $\div$ 189&--106 $\div$ 216&--121 $\div$ 231\\
$\Delta V_{ch}$ [km s$^{-1}$]&0.37&0.21&0.29\\
$<\sigma_{\rm rms}>$ [K] &0.09&0.42&0.93\\
\noalign{\smallskip}
\tableline
\end{tabular}
}
\end{center}
\end{table}
\begin{figure}
\includegraphics[scale=0.6]{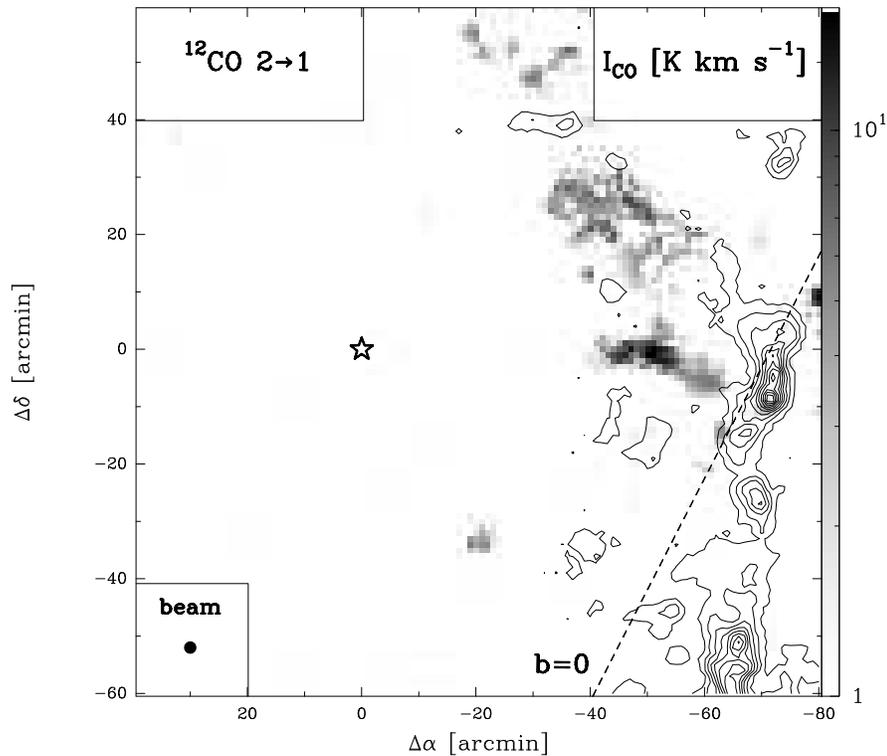}
\caption{The intensity map of CO $J$ = 2$\to$1 integrated over two
  velocity ranges: 18--32 km s$^{-1}$ (contours) and 44--57 km
  s$^{-1}$ (grey scale). Contours are plotted from 2.7 to 64.1 K~km~s$^{-1}$ by
  6.8  K~km~s$^{-1}$ (4 to 94\% by 10\% of the maximum). The V838~Mon
  position is marked a star-like symbol. The Galactic equator is
  shown as a dashed line.}
\end{figure}        

Sparks et al.~(2006, this issue) have concluded from polarimetric observations
of the light echo that the distance to V838~Mon is
$\sim 5.9$~kpc. This value is close to the distance of the outermost spiral arm in
the star direction i.e. Norma--Cygnus Arm. Thus
V838~Mon is likely to be situated {\it in} or at least {\it close to} this spiral
arm. This is consistent with the radial velocity of the star of
$V_{\rm lsr}=54.3$ km s$^{-1}$ \citep{deguchi}, the existence of the
B3 V companion and the proposed 
membership of V838~Mon to a young cluster of B-type stars 
(Bond et al.~2006, this issue). 
%%%%%%%%%%%%%%%%%%%%%%%%%%%%%%%%%%%%%%%%%%%%%%%%%%%%%%%%%%%%%%%%%%%%%%
\subsection{Observations in CO \boldmath $J$ = 1$\to$0}
Principle technical details of our observations in CO (1--0) are
presented in Table~1. Average 3$\sigma_{\rm rms}$ level for these
observations is of $\sim$0.3~K, which gives a better sensitivity than that
of the Galactic survey by \citet{dame} in the same transition.   

Significant CO~(1--0) emission is seen only for points west of the
  star position, in the area where we have also found emission in CO
  (2--1) and (3--2) (see \S3.1). The emission is always
  related to molecular clouds \citep[what has already been noted
  by][]{tyl}. Among the points in the ring-like structure claimed in
\citet{loon} we have found
a clear emission only at an offset $\Delta
  \alpha$=12', $\Delta \delta$=1' (see upper and right axes in Fig.~1,
  the offset is given in respect to the star position). The emission
  is weak, $T_A^* \approx$0.6~K, and narrow, $\Delta
  V$=1.9~km~s$^{-1}$. It is at a radial velocity of $V_{\rm
  lsr}$=49.2~km~s$^{-1}$, only $\sim$5~km~s$^{-1}$ different from the
  velocity of V838~Mon (54.3~km~s$^{-1}$). 

Our observations in CO~(1--0) show that the shell-like structure suggested by
  \citet{loon} does not exist. Detection of the emission at 12' from the
  star (20~pc at 6~kpc) suggests that there can be a
  weak extended molecular emission near V838~Mon, below the
  sensitivity of our observations. Observations with a better
  sensitivity are needed to verify if this molecular matter can be
  related to the echoing matter.                 

\section{Observations at the position of V838 Mon}
V838 Mon itself was observed in the three lowest rotational transitions of CO
with longer integration times. Observational details are summarized in
Table~2.  
 
In the CO (1--0) transition the V838~Mon position was observed only once. We did
not find any emission stronger than $0.2$~K ($3\sigma_{\rm rms}$). We also
observed 8 positions around the star, i.e. points with offsets of $\pm 1$', 
with shorter integration times. In  all these
positions we did not detect any emission higher than $0.6$~K
($3\sigma_{\rm rms}$). The area covered by these 9 points 
corresponds to that of the light echo in the fall of
2005. Thus the echoing matter is not seen at the sensitivity of our observations.  

In CO (2--1) and (3--2) we have spectra obtained in April and December
2005. In the spring spectra we have found emission only in the CO~(2--1)
transition at a radial velocity of $V_{\rm lsr}=53.4$~km~s$^{-1}$, almost
the same as the velocity of the SiO maser emission \citep{deguchi}. 
In the spectra obtained in December
with much longer integration times we have found two components
in both transitions. Apart from the already known component at $V_{\rm
  lsr}=53.4$~km~s$^{-1}$, an emission is also seen at a radial velocity of
$V_{\rm lsr}=-11$~km~s$^{-1}$. The line intensities, widths
and peak velocities are presented in Table~3. All the lines are very
narrow. The component at $\sim 53$~km~s$^{-1}$ is stronger in the CO
(2--1) transition, while that at $\sim -11$~km~s$^{-1}$ is
stronger in CO (3--2). When one compares the line intensities in April and
December it looks like the strength of the component at $\sim
53$~km~s$^{-1}$ decreased. However,
the observed change is close to the measurment uncertainty so this finding
is not conclusive.

The beam widths of our observations in CO (2--1) and (3--2) are rather
large (130'' and 82'', respectively). Therefore we cannot say whether the
emission detected at the V838 Mon position comes from the closest stellar
vicinity or from regions which are located at a few tens of arcmin from
the star. It should be noted that a search for molecular emission
(including CO and SiO maser) in 2003 gave negative result \citep{rushton}. 
The detected
emission in the CO lines may indicate that a molecular
activity has started close to V838~Mon, but its origin is unclear. 
Further
observations with a better angular resolution would be of particular interest
in this point.
\begin{table}
\caption{The technical description of the observations at
  V838 Mon position in the CO lines in 2005.}
\smallskip
\begin{center}
{\small
\begin{tabular}{ccccccc}
\tableline
\noalign{\smallskip}
date&line&$T_{\rm sys}$&$\Delta \nu_{ch}$&$\Delta V_{ch}$&velocity range&$\sigma_{\rm rms}$\\
&&[K]&[kHz]&[km~s$^{-1}$]&[km~s$^{-1}$]&[K]\\
\noalign{\smallskip}
\tableline
\noalign{\smallskip}
04 Apr.&2--1&168&165.4&0.21&--106.4 $\div$ 171.4&0.043\\
       &3--2&273&339.1&0.29&--121.4 $\div$ 186.4&0.083\\
       &    &   &     &    &                    &     \\
02 Nov.&1--0&245&142.1&0.37&--139.5 $\div$ 188.7&0.072\\       
       &    &   &     &    &                    &     \\
26 Dec.&2--1&310&165.4&0.21&--206.8 $\div$ 233.5&0.014\\
       &3--2&418&339.3&0.29&--289.5 $\div$ 250.4&0.022\\
\tableline
\noalign{\smallskip}
\end{tabular}
}
\end{center}
\end{table}
%%%%%%%%%%%%%%%%%%%%%%%%%%%%%%%%%%%%%%%%%%%%%%%%%%%%%%%%%5
\begin{table}
\caption{Spectral chracteristics of the CO lines found in the observations at V838 Mon position in 2005.}
\smallskip
\begin{center}
{\small
\begin{tabular}{ccrccc}
\tableline
\noalign{\smallskip}
date&line&\multicolumn{1}{c}{$V_{\rm LSR}$}&\multicolumn{1}{c}{Peak $T_{\rm mb}$}&FWHM&Int. intensity\\
    &    &[km~s$^{-1}$]                     &\multicolumn{1}{c}{[K]}                &[km~s$^{-1}$]&[K km~s$^{-1}$]\\
\noalign{\smallskip}
\tableline
\noalign{\smallskip}
04 Apr.&2--1&  53.3 &0.553&1.174&0.691\\
26 Dec.&2--1& --11.1&0.099&1.075&0.114\\
       &    &  53.3 &0.323&1.179&0.406\\
       &3--2& --11.0&0.188&1.148&0.230\\
       &    &	53.2&0.156&1.500&0.190\\
\tableline
\noalign{\smallskip}
\end{tabular}
}
\end{center}
\end{table}
\acknowledgements We wish to thank all the staff at Delingha, the
milli\-me\-ter-wave radio observatory of Purple Mountain Observatory for
the observations in CO (1--0). 

The KOSMA 3 m telescope is operated by
the K\"{o}lner Observatorium f\"{u}r SubMillimeter Astronomie of the
I.~Physikalisches Institut, Universit\"{a}t zu K\"{o}ln in
collaboration with the Radioastronomisches Institut, Universit\"{a}t
Bonn. 

The research reported in this paper
has partly been supported by the Polish State Committee for Scientific
Research through grants no. 2 P03D 002 25 and 2 P03D 017 25.

\end{document}